Superconductivity of FeSe$_{1-x}$Te$_x$ Thin Films Grown by Pulsed Laser Deposition

Yoshinori Imai[1, 3, *], Ryo Tanaka[1, 3], Takanori Akiike[1, 3], Masafumi Hanawa[2, 3], Ichiro Tsukada[2, 3], and Atsutaka Maeda[1, 3]

[1] Department of Basic Science, the University of Tokyo, 3-8-1 Komaba, Meguro-ku, Tokyo 153-8902, Japan

[2] Central Research Institute of Electric Power Industry, 2-6-1 Nagasaka, Yokosuka, Kanagawa 240-0196, Japan

[3] Transformative Research-Project on Iron Pnictides (TRIP), JST, Japan.

Abstract

FeSe$_{1-x}$Te$_x$ thin films with PbO-type structure are successfully grown on MgO(100) and LaSrAlO$_4$(001) substrates from FeSe$_{1-x}$Te$_{x+d}$ polycrystalline targets by pulsed-laser deposition. The film deposited on the MgO substrate (film thickness ~ 63 nm) shows superconductivity at 6.7 K (onset) and 5.5 K (zero resistivity). On the other hand, the film deposited on the LaSrAlO$_4$ substrate (film thickness ~ 52 nm) exhibits only the onset of superconducting transition at 2.6 K, but does not show zero resistivity. This indicates the strong influence of epitaxial strain to superconducting properties of FeSe$_{1-x}$Te$_x$ thin films.

Footnotes:

(*) E-mail address: imai@maeda1.c.u-tokyo.ac.jp

Since the discovery of the superconductivity in F-doped LaFeAsO,[1] considerable studies on the iron-based superconductors have been done and have attracted much attention. FeSe with superconducting transition temperature, $T_c$, of 8 K[2] is one of the iron-based superconductors and has the simplest tetragonal PbO-type structure. Nevertheless, it is expected that FeSe is electrically quasi two-dimensional as other iron-based superconductors. It has turned out that measurements of the complex conductivity in the microwave and terahertz region provide useful information for cuprate superconductors which are well known to be quasi two-dimensional superconductors with the $CuO_2$ plane.[3-5] Thus, similar studies about a high-frequency conductivity are extremely interesting in terms of mechanism of superconductivity in this system. These measurements require high-quality single-crystalline thin films that are prepared on a substrate with a small dielectric constant. So far, successful growth of $FeSe_{1-x}Te_x$ films has been reported in literature only by a pulsed laser deposition (PLD) method.[6] The growth condition and detailed method, however, have not been optimized yet. In this paper, we report PLD growth of $FeSe_{1-x}Te_x$ superconductors using the substrates applicable to high-frequency measurements; MgO and $LaSrAlO_4$ (LSAO) single crystals. The $a$-axis length of MgO is 4.21 Å and is far longer than that of $FeSe_{1-x}Te_x$ ($a$ = 3.76-3.82 Å),[7] while $LaSrAlO_4$ ($a$ = 3.76 Å) has the better lattice matching to $FeSe_{1-x}Te_x$. We investigated these two substrates comparatively and found that a worse lattice matching gives better superconductivity, contrary to the simple expectation before starting the experiments.

Polycrystalline pellets with the nominal composition of $FeSe_{0.5}Te_{0.75}$ were used as the target, which were synthesized as follows. First, mixed powders of the starting materials, Fe, Se and Te with the molar ratio of 1 : 0.5 : 0.5, were pelletized and heated in the evacuated quartz tube at 953 K for 12 h. Next, reground powders of the reacted pellets and additional Te with molar ratio of 1 : 0.25 were pelletized again and heated in the evacuated quartz tube at 953 K for 2 h. Thin films were grown under vacuum (~ $10^{-5}$ Torr) by a pulsed laser deposition method using KrF laser (wavelength: 248 nm, repetition rate, $f$: 1-10 Hz). We use a metal mask to prepare $FeSe_{1-x}Te_{x+d}$ films in a six-terminal shape for transport measurements. Substrate temperature, $T_s$, was set between 573 and 673 K. We kept the film thickness below 100 nm because of the possible application to the transmission terahertz spectroscopy. The crystal structure and the orientation of films were characterized by x-ray diffraction (XRD) using Cu $K_\alpha$ radiation at room temperature. Electrical resistivity measurements were carried out by a four-terminal method from 2 to 300 K with magnetic fields up to 13 T applied perpendicular to the film surface. Hall effect was also measured for the same sample with sweeping field between −2 T to 2 T.

Figure 1 shows the XRD pattern of the film deposited at $T_s$ = 573 K with $f$ = 10 Hz on MgO. Only the 00$l$ reflections of a tetragonal PbO-structure are observed, which indicates that the film is preferentially oriented along the $c$-axis. The deposition conditions, film thicknesses, and $c$-axis lattice parameters of several films are summarized in Table I. The $c$-axis lengths of films D and E are smaller than those of the other films, which are comparable with that of FeSe$_{0.5}$Te$_{0.5}$ bulk sample ($c$ = 5.89 Å).[7] According to the dependence of $c$-axis length on Te content in FeSe$_{1-x}$Te$_x$ bulk samples,[7] the Te contents in films D and E are likely to be less than those of the other films. This result suggests that the Te content in the film depends on $T_s$ rather than the kinds of substrate material. To clarify it, it is necessary to perform a composition analysis such as Electron Probe Micro-Analysis (EPMA).

The temperature dependence of the electrical resistivity of the five films in Table I is shown in Fig. 2. Films A and B grown on MgO exhibit superconductivity with $T_c^{onset}$ (defined as the temperature where the electrical resistivity deviated from the normal state behavior) of 5.0 K and 6.7 K and $T_c^{zero}$ (defined as the temperature where the electrical resistivity becomes zero) of 3.7 K and 5.5 K, respectively. Both $T_c^{zero}$ and $T_c^{onset}$ increase with the film thickness, which is qualitatively consistent with the tendency that was reported before.[6] We also emphasize that zero resistance shows up in films that are less than 100 nm in thickness, in contrast to the results reported in ref. 6. We speculate that the better results in films A and B than the previous report[6] is because of less deficiency of Te caused by the use of the polycrystalline targets containing excess Te. Indeed, the *c*-axis lengths of these films are close to those of bulk FeSe$_{0.5}$Te$_{0.5}$ samples, which supports our speculation. As opposed to films A and B grown at $T_s$ = 573 K, superconductivity is completely suppressed above 2 K in film D grown on MgO at $T_s$ = 673 K. This is probably because the high substrate temperature enhances the deficiencies of Se and/or Te that have a high vapor pressure. As for the films prepared on LSAO, the electrical resistivity of film C slightly drops at $T_c^{onset}$ ~ 2.6 K but does not become zero completely, and even the onset of superconductivity is not observed in film E. This is probably due to the lack of continuous superconducting current path through the entire films. This result is surprising at first sight, considering that the film-to-substrate lattice mismatch is far smaller for LSAO than for MgO. We consider the reason for the surprising result as follows. First, concerning of the films grown on MgO, the mismatch is so large that the lattice strain is relaxed within a few layers from the substrate surface. The resultant films have the lattice parameters which are similar to those of bulk samples. This probably leads to the appearance of superconductivity with zero resistance even in films less than 100 nm in thickness. On the other hand, as for the films grown on LSAO, we assume that the film is grown without relaxing the lattice strain. Then, the insufficient superconductivity property of films grown on LSAO is due to the influence of the epitaxial strain caused by the substrate. Thus, the result suggests that the superconductivity of FeSe$_{1-x}$Te$_x$ is strongly influenced by the epitaxial strain. A further systematic investigation of the delicate epitaxial-strain effect is necessary to find a route to higher $T_c$ in this system, which is the subject of subsequent studies.

The effect of applied magnetic field on the electrical resistivity in film B was shown in Fig. 3. $T_c$ decreases linearly with increasing magnetic field. The upper critical field, $H_{c2}$, which is defined as a field where the resistance becomes half the value of normal-state resistance, is plotted in the inset of Fig. 3 as a function of temperature near $T_c$. $H_{c2}$ extrapolated to $T = 0$ K,[8] namely $H_{c2}(0)$, is estimated to be 28.7 T. That gives a Ginzburg-Landau coherence length $\xi_a \sim 33.8$ Å.

Shown in Fig. 4 are the data of the temperature dependence of Hall coefficient ($R_H$) for film B. The temperature dependence of $R_H$ is complicated, being consistent with the expectation that both hole- and electron-like bands contribute to the electric transport in FeSe$_{1-x}$Te$_x$.[9,10] One of the characteristics is that the temperature dependence of $R_H$ looks highly correlated to $\rho(T)$. This strongly suggests that the complicated behavior of $\rho(T)$ is a direct consequence of the change of Fermi-surface topology such as a magnetic instability. We will investigate the Te contents (thus carrier doping) dependence of Hall effect in more detail.

In conclusion, we successfully grow the $FeSe_{1-x}Te_x$ superconducting films by a PLD method. The film deposited on the MgO(100) substrate with film thickness ~ 63 nm has shown superconductivity at 6.7 K (onset) and 5.5 K (zero resistivity). The appearance of superconductivity in the very thin film results from the use of the target containing excess Te. The film deposited on the $LaSrAlO_4$ substrate, on the other hand, shows only the onset of superconducting transition at 2.6 K, but not zero resistivity. The present results indicate the presence of a strong epitaxial-strain effect between $FeSe_{1-x}Te_x$ and $LaSrAlO_4$, though the reason has not been clarified why the substrate with closer lattice mismatch gives worse influence on the superconductivity of $FeSe_{1-x}Te_x$ thin films.

Figure 1. X-ray diffraction pattern of film B grown at $T_s$ = 573 K with $f$ = 10 Hz on MgO.  The asterisks represent the peaks resulting from the substrate.

Figure 2. The temperature dependence of the electrical resistivity for the films shown in Table I.  The inset shows the electrical resistivity near $T_c$.

Figure 3. The temperature dependence of the electrical resistivity for film B in applied magnetic field.  The upper critical field $H_{c2}$ which is defined as a field where the resistance becomes half the value of normal-state resistance is plotted in inset as a function of temperature near $T_c$.

Figure 4. The temperature dependence of the Hall coefficient for film B.

Table I. The specifications of the grown films.

|   | $T_s$ (K) | $f$ (Hz) | Substrate | Film thickness (nm) | $c$-axis length (Å) |
|---|---|---|---|---|---|
| A | 573 | 1 | MgO | 29 | 5.83 |
| B | 573 | 10 | MgO | 63 | 5.86 |
| C | 573 | 10 | LSAO | 52 | 5.84 |
| D | 673 | 10 | MgO | 65 | 5.77 |
| E | 673 | 10 | LSAO | 50 | 5.78 |

Figure 1

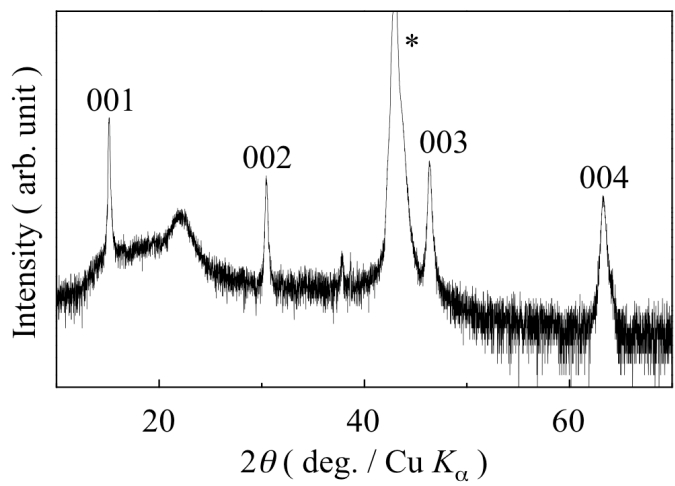

Figure 2

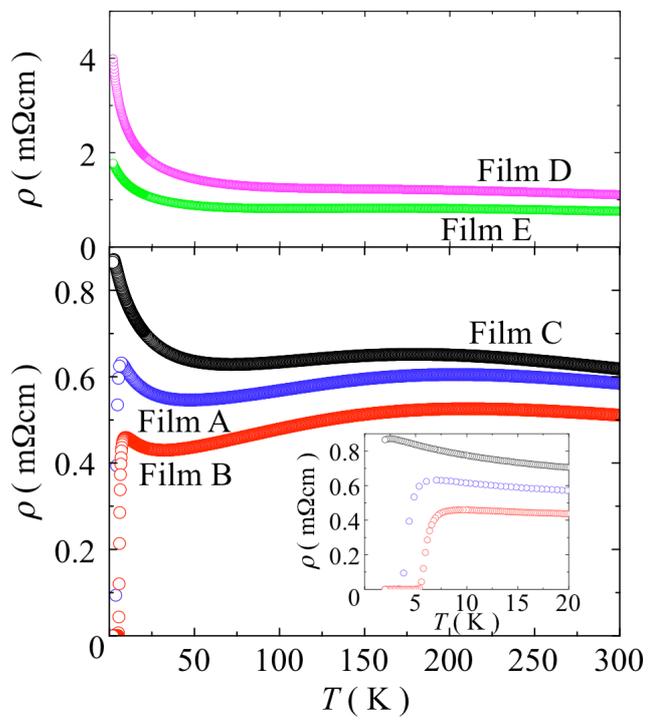

Figure 3

Figure 4

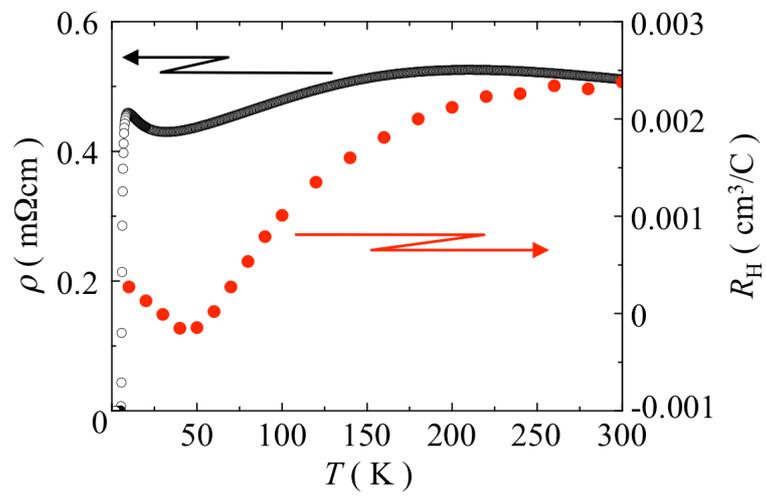